\documentclass[10pt]{wlscirep}
\usepackage{hyperref}
\usepackage[english]{babel}

\title{Operating Nanobeams in a Quantum Fluid}

\author{D.\,I.~Bradley}
\author{R.~George}
\author{A.\,M.~Gu\'{e}nault}
\author{R.\,P.~Haley}
\author{S.~Kafanov$^*$}
\author{M.\,T.~Noble}
\author{Yu.\,A.~Pashkin}
\author{G.\,R.~Pickett}
\author{M.~Poole}
\author{J.\,R.~Prance}
\author{M.~Sarsby}
\author{R.~Schanen}
\author{V.~Tsepelin$^\dagger$}
\author{T.~Wilcox}
\author{D.\,E.~Zmeev}

\affil{Department of Physics, Lancaster University, Lancaster, LA1 4YB, United Kingdom \\
[$*$] sergey.kafanov@lancaster.ac.uk\\
[$\dagger$] v.tsepelin@lancaster.ac.uk}

\begin{abstract}
Microelectromechanical (MEMS) and nanoelectromechanical systems (NEMS) are ideal candidates for exploring quantum fluids, since they can be manufactured reproducibly, cover the frequency range from hundreds of kilohertz up to gigahertz and usually have very low power dissipation. Their small size offers the possibility of probing the superfluid on scales comparable to, and below, the coherence length.  That said, there have been hitherto no successful measurements of NEMS resonators in the liquid phases of helium.  Here we report the operation of doubly-clamped aluminium nanobeams in superfluid $^4$He at temperatures spanning the superfluid transition. The devices are shown to be very sensitive detectors of the superfluid density and the normal fluid damping. However, a further and very important outcome of this work is the knowledge that now we have demonstrated that these devices can be successfully operated in superfluid $^4$He, it is straightforward to apply them in superfluid $^3$He which can be routinely cooled to below 100\,$\mu$K. This brings us into the regime where nanomechanical devices operating at a few MHz frequencies may enter their mechanical quantum ground state.
\end{abstract}

\begin{document}
\maketitle
\thispagestyle{empty}

\section*{Introduction}
\label{sec: Introduction}
The relentless drive to reduce the size of electronic components, coupled with the continuous progress in fabrication technology,  has given us the capability of creating complex structures on the micron and submicron scale.  In consequence MEMS and NEMS are becoming common research tools in the areas of mass and force sensing \cite{Nature.446.1066.Burg, NatNanotech.7.301.Chaste}, atomic force microscopy \cite{NatureNano.4.1748.Wilson}, nanofluidics \cite{Nanolett.15.8070.Kara} and quantum behaviour of macroscopic mechanical oscillators \cite{Nature.464.697.OConnell, Science.349.952.Wollman}.

In this paper we address the efforts to develop new MEMS and NEMS devices, or to adapt those already available, for use in liquids at cryogenic temperatures \cite{Nanotechnology.11.165.Kraus, J.Low.Temp.Phys.183.284.Defoort, PhysRevLett.113.136101.Defoort, Rev.Sci.Instr.84.025003.Gonzales, Nat.Phys.3813.Harris}. There are two very major aims here. First, a micro- or nanoscale resonator immersed in a pure superfluid would immediately allow us to probe its properties in the completely novel regimes made accessible when the device dimensions become comparable to the coherence length of the liquid.

Secondly, however, there is a further, wider motivation. Accessing the quantum regime for mechanical systems should lead to the realm of quantum mechanical systems where bulk mechanical objects consisting of billions of atoms are governed by the laws of quantum physics \cite{PhysToday.58.7.36.Schwab, PhysRep.511.273.Poot}. At the lowest temperatures and for very small displacements, immersing a NEMS device in a superfluid (which is effectively a ``mechanical vacuum” but still has good thermal properties) may enable the cooling of devices\cite{Nat.Commun.10455.Bradley} down to the mechanical-ground-state level using ``brute force'' cooling, which so far has been accomplished using much higher frequency systems\cite{Nature.464.697.OConnell}. As a first step we describe here the use of a 1\,MHz resonator in superfluid $^4$He down to $\sim$1\,K.  For such a resonator the quantum ground-state regime would be reached at a temperature of $\sim$100\,$\mu$K. We routinely achieve such temperatures in superfluid $^3$He cooled by the nuclear demagnetization of copper. Thus the devices described here, if immersed in superfluid $^3$He as the thermal contact agent, should readily reach the quantum regime.
	
Historically, mechanical resonators have been widely used to probe bulk properties of $^3$He and $^4$He superfluids \cite{PhysicaB.197.390.Pickett, JPC.4.129.Black}, such as their transport properties, to create and detect the presence of topological structures/defects such as quantum vortices \cite{PhysRevLett.82.4831.Stalp, Nature.4.46.Bradley, NatPhys.7.473.Bradley}, and to study the pairing configurations of the various $^3$He phases \cite{Nat.Phys.3813.Bradley}. The most long-standing and commonly used mechanical resonator is the vibrating wire \cite{JLowTempPhys.62.511.Guenault}, typically comprising a few millimetres of superconducting wire, with diameters in the range of $1-100$ micrometres, and with resonance frequencies of up to few kilohertz. A vibrating wire immersed in superfluid can act as both a generator and detector of excitations and as a secondary thermometer down to millikelvin temperatures in $^4$He \cite{PhysRevLett.100.045301.Yano} and tens of $\mu$K in $^3$He \cite{Physica.126B.260.Guenault}. During the past decade a range of newer devices have been employed  to study bulk quantum fluids: miniature piezoelectric quartz tuning fork oscillators with frequencies up to hundreds of kilohertz \cite{PhysRevB.85.014501.Bradley, PhysRevB.89.014515.Ahlstrom}, aluminium-coated silicon ``goal-post'' shaped micro-electro-mechanical systems \cite{J.Low.Temp.Phys.183.284.Defoort} and comb-drive electrostatic MEMS \cite{Rev.Sci.Instr.84.025003.Gonzales}.  Nevertheless, superconducting wire of a few microns diameter has been the most convenient probe to measure the lowest temperatures inside the superfluid $^3$He via the damping and frequency shift of the resonance curve. It has even been proposed that vibrating wire bolometry might provide ultra-sensitive low temperature dark-matter particle detectors\cite{Pickett1988,PhysRevLett.75.1887.Bradley}.

Prior to the current work, NEMS devices have only been successfully used at cryogenic temperatures to probe vapour properties \cite{PhysRevLett.113.136101.Defoort, Nanotechnology.11.165.Kraus}, since bulk cryogenic liquids were perceived to be too difficult to study. Recently, the properties of a superfluid film have been measured by using an opto-mechanical oscillator \cite{Nat.Phys.3813.Harris}. The work presented here constitutes the first successful measurement of bulk superfluid properties using NEMS resonators and provides a significant step towards both building routine superfluid probes and the ambitious goal of cooling a NEMS beam to its quantum-mechanical ground state using a quantum fluid as the cooling medium.

\section*{Results}
\subsection*{Beam Characteristics in Vacuum}
\label{sec: VacuumProp}
\begin{figure}
\includegraphics[width=\linewidth]{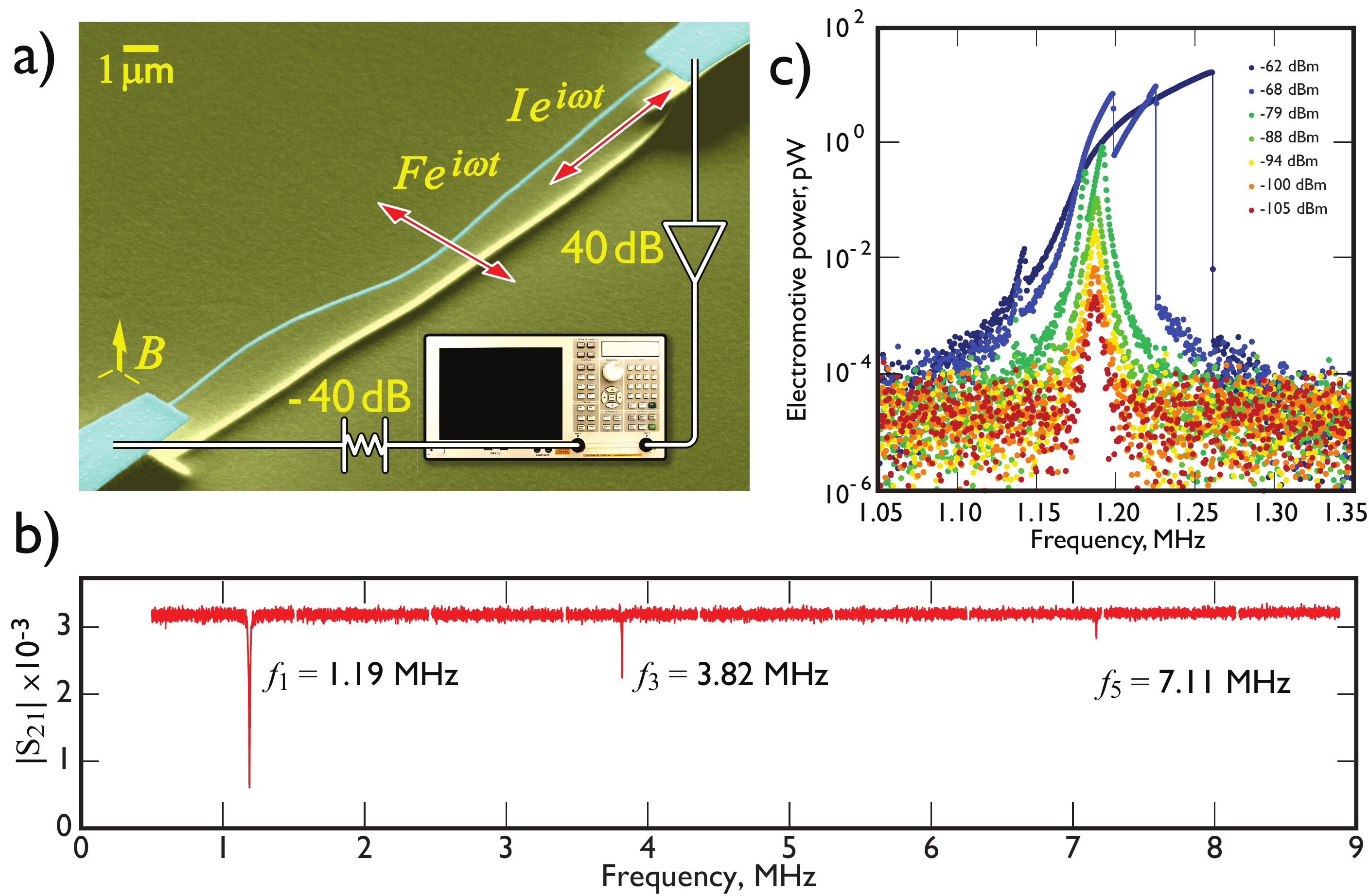}
\caption{The behaviour of the aluminium beams in vacuum. \textbf{(a)} A micrograph of a typical doubly-clamped aluminium beam as used in experiments. Despite an obvious compression of the resonator at room temperature, our beams have a tensile stress at cryogenic temperatures due to the significant differential thermal contraction of the silicon substrate and the deposited aluminium film. The resonator shown here has a nominal length of 15\,$\mu$m with a cross-sectional dimensions of 0.1\,$\mu$m\,$\times$\,0.1\,$\mu$m, with an expected resonance frequency at cryogenic temperatures of 8.5\,MHz. The beam is placed in a perpendicular magnetic field and connected in the microwave circuit, shown schematically in the figure. Transmission measurements are performed with a network analyser. The microwave drive from the network analyser is attenuated by -40\,dB with an attenuator located at 4.2\,K; the microwave signal transmitted through the sample is amplified by a 40\,dB amplifier at room-temperature. \textbf{(b)} A typical microwave transmission measurements of our $50\,\mathrm{\mu m}$ beam in vacuum at 4.2\,K in a 5\,T magnetic field. A wide frequency sweep at -90\,dBm of applied power clearly shows the first three odd harmonics of the beam. The quality factors of all three harmonics are approximately 10$^3$. \textbf{(c)} The frequency characteristics of electromotive power generated by a 50\,$\mu$m aluminium beam measured over a range of applied powers. The beam drive power in dBm is shown in the figure. The typical frequency response shows a Lorentzian shape at low driving power. However, all measured resonators demonstrate a Duffing response at drive powers above -86\,dBm. This particular beam has an additional parasitic resonance becoming visible above -88\,dBm, probably the second ``near-degenerate’’ perpendicular beam mode.}
\label{fig: VacuumMeasurements}
\end{figure}

The aluminium NEMS beams used in our experiments were formed lithographically on a silicon substrate and after fabrication were measured using a standard microwave magnetomotive technique as shown in figure \ref{fig: VacuumMeasurements}(a). The beam is driven by an oscillating current which provides the necessary lateral Lorentz force generated in the ambient magnetic field (typically 5\,T). The details of manufacturing and measurement setup are described in the Methods section.

Initial characterisations of our beams were performed in vacuum at a temperature of 4.2\,K. An example of the frequency response of a 50\,$\mu$m beam in vacuum is shown in figure \ref{fig: VacuumMeasurements}(b). Resonance peaks were detected at frequencies of 1.19\,MHz, 3.82\,MHz and 7.11\,MHz corresponding to the first three odd harmonics of the transverse flexural oscillations. The measured resonance frequencies can be compared to the expected values for a doubly-clamped beam of rectangular cross-section given by \cite{Bao}:
\begin{equation}
\label{eqn: BeamFreq}
f_n=\frac{k_n^2}{\pi\sqrt{48}}\frac{w}{l^2}\sqrt{\frac{E}{\rho_\mathrm{beam}}}\sqrt{1+\gamma_n\left(\frac{l}{w}\right)^2\frac{\Delta l}{l}},
\end{equation}
where $w$ and $l$ represent  the width and the length of the beam, respectively. The coefficients $k_n$ and $\gamma_n$ have different values depending on the eigenmode of the resonance: \cite{Bao} $k_1=4.7300$, $\gamma_1=0.2949$, $k_2=7.8532$, $\gamma_2=0.1453$, and $k_{n\geqslant 3} = \pi(n + 1/2)$, $\gamma_{n\geqslant 3}=12(k_n-2)/k_n^3$.  Despite a weak granularity of our aluminium films, we assume that the Young's modulus and density of the beams are close to the bulk values of $E = 70$\,GPa and $\rho_\mathrm{beam}=2700$\,kg\,m$^{-3}$ respectively. The ratio $\Delta l/l$ describes the tensile strain of the resonators, which is expected to be substantial at cryogenic temperatures owing to the mismatch in the thermal expansion coefficients of the silicon substrate and aluminium film. Simple calculation shows that the unstressed 50\,$\mu$m long NEMS beam, with an approximately square cross-section of 0.1\,$\mu$m\,$\times$\,0.1\,$\mu$m, should have a fundamental resonance at 209\,kHz. The analysis of measured first, third and fifth eigenmodes shows that the actual average width, $w$, of our beams is about 0.18\,$\mu$m and the value of the tensile strain, $\Delta l/l\approx4.3\cdot10^{-4}$, is an order of magnitude lower than that measured for shorter aluminium resonators \cite{Appl.Phys.Lett.92.043112.Li, Nano.Lett.10.4884.Sulkko}. We note that, due to an extremely high aspect ratio ($l/w\gtrsim10^2$), our resonators have a significant compression stress at room temperature, which is clearly observable as bending on the micrograph of our typical beam (see figure \ref{fig: VacuumMeasurements}(a)), arising from the forced match of  the substrate and deposited lattices at the time of deposition. Such obvious compression stress is not seen in shorter samples in other experiments \cite{Appl.Phys.Lett.92.043112.Li, Nano.Lett.10.4884.Sulkko}.		

The power dependence of the frequency response of the 50\,$\mu$m aluminium beam near the first harmonic is presented in figure~\ref{fig: VacuumMeasurements}(c). At low excitations the beam exhibits a linear response and has a quality factor ($Q$-factor) on the order of 10$^3$. Similar $Q$-factors are observed for higher harmonics and other beams in vacuum at 4.2\,K. At high excitations the nonlinearities stiffen the beam and the resonance peak position shifts to a higher frequency as expected for the Duffing oscillator \cite{MicrosystemTechnologies.1432.1.Tajaddodianfar}. In addition to Duffing-like non-linearity, above intermediate applied power (-88\,dBm) a second resonance becomes apparent which we believe is the lower-level excitation of the second of the two near-degenerate flexural modes of the almost square cross-section beam.
			
Since our ultimate goal is to probe liquid helium, it is instructive to deduce the force-velocity characteristics of the beam from the frequency response as a function of power to confirm the velocity below which the beam response remains linear. We have used the definition of quality factor $Q = f_1/\Delta f = \pi f_1 \, m_\mathrm{beam} v^2/P_{\max}$ to deduce the force $F_\mathrm{L}$ and velocity $v$ values, and applied them even when the resonance curve became non-Lorentzian:
\begin{equation}
v\approx\sqrt{\dfrac{P_{\max}}{\pi m_\mathrm{beam} \Delta f}}; 
\qquad
F_\mathrm{L}\approx\sqrt{\pi m_\mathrm{beam} P_{\max} \Delta f}.
\label{eqn: ForceVelocity}
\end{equation}
Here $m_\mathrm{beam}=\rho_{\mathrm{beam}}V$ is the mass of the beam, $\Delta f$ is the width of the resonance and $P_{\max} = F_\mathrm{L}v$ is the maximum observable power amplitude. We conclude that the beam behaves linearly up to a velocity amplitude of 0.1\,m\,s$^{-1}$ corresponding to a displacement of 13\,nm, above which it behaved non-linearly. Therefore any non-linear behaviour observed in a liquid medium below that velocity should be attributed to the interaction of the beam with the fluid.

\subsection*{Beam response in liquid helium}
\label{sec: ProbingLiquid}

\begin{figure}
\includegraphics[width=0.9\linewidth]{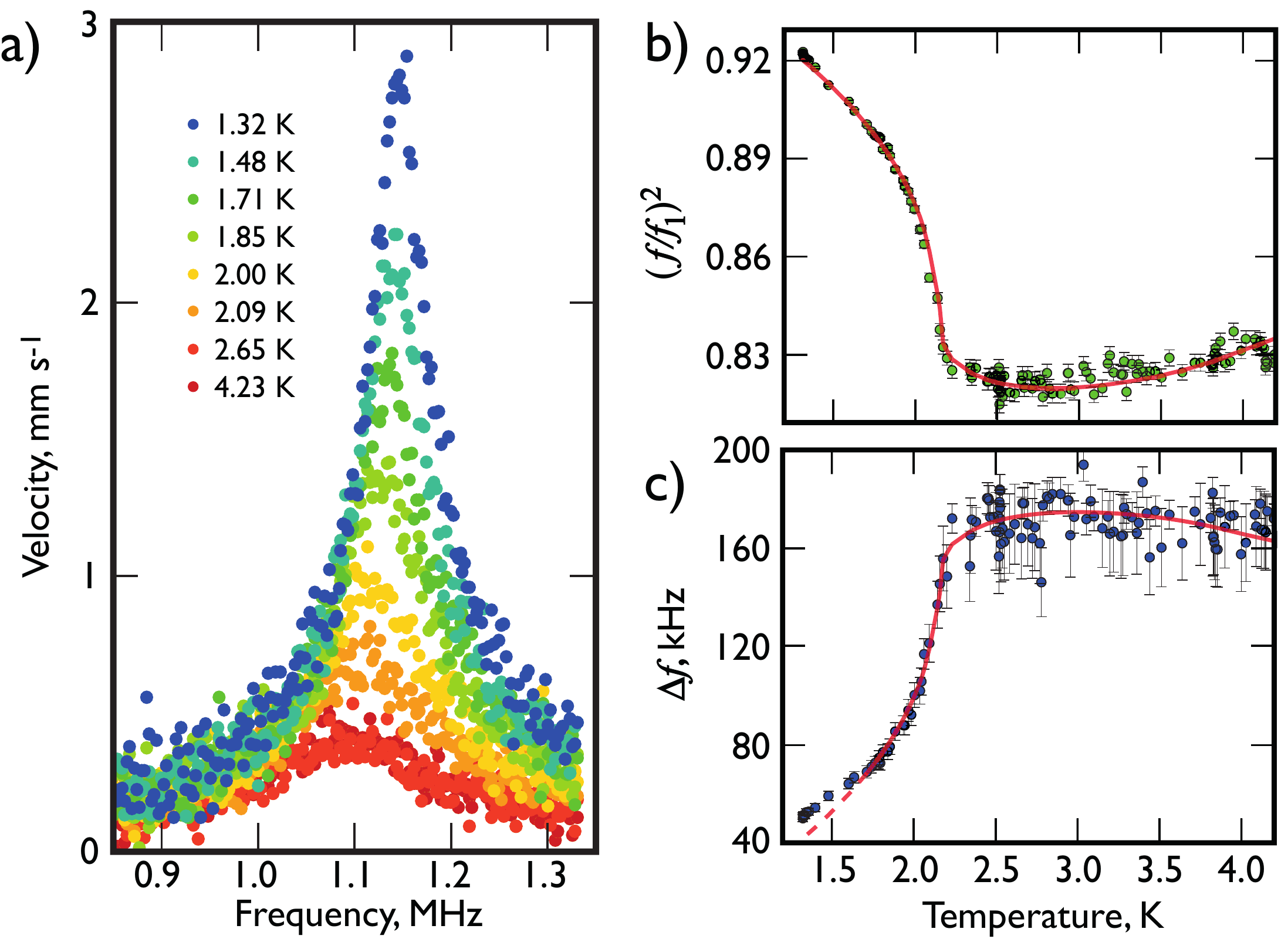}
\caption{The response of the beam in liquid $^4$He. \textbf{(a)} The velocity resonance curves of the 50\,$\mu$m beam oscillating in liquid $^4$He. The output power of the network analyser was maintained at a constant -50\,dBm with an additional -40\,dB external attenuation. The shape of the resonance is independent of temperature in the normal fluid above the transition temperature $T_\lambda=2.178$\,K. The decrease in density of the normal component of the liquid below $T_\lambda$ results in a decrease of the Stoke's drag, leading to a significant increase in the NEMS quality factor along with  an increase in the resonant frequency as less fluid is being dragged with the beam. \textbf{(b, c)} The temperature dependence of the resonance frequency and the resonance width of the 50\,$\mu$m nanomechanical resonator. Each data point of figures \textbf{(b)} and \textbf{(c)} was obtained from a complete resonance curve for the beam. The solid lines are the theoretical models given by equations~(\ref{eqn: FreqRatio}) and (\ref{eqn: DeltaF}) for the \textbf{(b)} and \textbf{(c)} respectively. We treated all three geometrical factors as fitting parameters and have found the following values: $\beta=1.18 \pm 0.02$, $B=1.19 \pm 0.01$ and $C=2.62\pm0.06$. The values obtained are close to the theoretical expectations and broadly agree with the values observed for vibrating wire resonators and tuning forks. The resonance width was fitted at temperature range between 4.2\,K and 1.7\,K, where the hydrodynamic model agrees well with the experimental data. The red dashed line highlights the difference between Stokes' model and experimental data below 1.7\,K.}
\label{fig: LiquidHe4}
\end{figure}

Figure~\ref{fig: LiquidHe4}(a) presents the temperature dependence of the frequency response of the beam around the fundamental resonance.  The resonance curves for all temperatures are very broad with widths of order 100\,kHz, \textit{i.e.} with quality factors of order 10. This guarantees that in-liquid measurements are completely dominated by liquid-induced dissipation since the corresponding intrinsic beam dissipation is negligible in comparison. The resonance frequency of the beam is also significantly lower ($\sim$10\%) than the vacuum value owing to the increase in effective mass of the beam arising from the induced flow of surrounding liquid helium. Helium flow can be thought of as having two components, the ``pure potential” backflow of liquid needed to allow the passage of the beam, and the additional liquid that moves because it is viscously ``clamped” to the beam.  	
		
The frequency sweeps in normal helium, between 4.2\,K and 2.178\,K, are almost indistinguishable and for clarity only representative data sets at 4.2\,K and 2.6\,K are plotted in the figure. As the temperature falls below the superfluid transition at 2.178\,K, the damping experienced by the beam decreases and the resonance peak shifts towards higher frequencies (since below the transition decreasing masses of liquid are dragged along by the wire). The observed increase of the beam velocity with falling temperature reflects the decreasing damping with falling normal fluid fraction. At the lowest measured temperature of 1.3\,K, where the helium has the highest superfluid fraction $\sim$95\%, the beam has the smallest observed resonance width with a corresponding maximum velocity of 2.8\,mm\,s$^{-1}$  (which we note is much lower than the velocity for the onset of any intrinsic nonlinear behaviour of the beam in vacuum).

For a quantitative description of the beam behaviour in liquid helium, we compare our results with the phenomenological two-fluid model for the superfluid. In the linear regime, at low beam velocities, we can treat our ``beam + liquid'' system as an externally-driven damped harmonic oscillator with a variable effective mass $\tilde{m}$ and a constant effective elastic constant $k$ \cite{J.Low.Temp.Phys.146.537.Blaauwgeers}. The ratio of resonance frequencies of the ``bare'' oscillator in vacuum,  $f_1=\sqrt{k/4\pi^2m_\mathrm{beam}}$, and our ``beam + liquid'' system, $f=\sqrt{k/4\pi^2\tilde{m}}$, can be written as:
\begin{equation}
\label{eqn: FreqRatio}
\left(\dfrac{f_1}{f}\right)^2=\dfrac{\tilde{m}}{m_\mathrm{beam}}=1 + \beta \dfrac{\rho}{\rho_\mathrm{beam}} + B\dfrac{\rho_\mathrm{n}}{\rho_\mathrm{beam}}\dfrac{S}{V}\sqrt{\frac{\eta}{\pi\rho_\mathrm{n} f}}.
\end{equation}
The effective mass $\tilde{m}$ of NEMS beam with volume $V$ and surface area $S$ immersed in a fluid with the total density $\rho$, the normal component density $\rho_\mathrm{n}$ and viscosity $\eta$ consists of a self mass of the resonator $m_\mathrm{beam}$ and two additional contributions associated with: (\textit{i}) a mass of fluid proportional to the volume of the oscillating body $m_\rho\propto\rho V$; (\textit{ii}) a mass of fluid in a layer of thickness the viscous penetration depth $\delta=\sqrt{\eta/(\pi \rho_\mathrm{n} f)}$ dragged along by the oscillating motion $m_\eta\propto \rho_\mathrm{n} S\delta$. Coefficients $B$ and $\beta$ are geometry-dependent parameters. The theoretical predicted values in the case of an infinitely long rectangular cross-section beam with height $h$ and width $w$ moving perpendicular to the width are  $\beta = (\pi/4)h/w$ and $B=1$ \cite{JAppPhys.84.64.Sader}.
			
The damping experienced by our system includes any intrinsic damping such as inter-crystalline friction within the material of the resonator itself, and external forces such as Stokes' drag which describes the interactions with the surrounding fluid. At a temperature of 4.2\,K this latter contribution dominates even for a beam oscillating in a rarefied helium gas at pressures as low as $\sim$10\,Pa. The Stokes' drag force is proportional to the velocity of the oscillating body. The exact calculation of the proportionality coefficient requires the full solution of the flow field around the oscillating body and in the high-frequency limit is given by $CS\sqrt{\pi\rho_\mathrm{n}\eta f}$, where $C$ is a geometrical constant independent of the fluid. For an infinitely long cylinder oscillating perpendicular to its axis, the value of this constant is \cite{J.Low.Temp.Phys.146.537.Blaauwgeers} $C=2$. For our 50\,$\mu$m beam with a resonance frequency of 1.2\,MHz the viscous penetration depth is 80\,nm at 4.2\,K (150\,nm at 1.5\,K) and is comparable to the beam thickness, which means that the high-frequency approximation used here is marginal. That said, treating the cross-over from high-frequency to steady-flow behaviour even for a cylinder is a formidable (non-trivial) task even in classical hydrodynamics\cite{LandauLifshitz:FluidMechanics} and a full treatment would not significantly benefit the description and core understanding of our main results. The distance of the beam to the substrate at 2\,$\mu$m is far enough away not to influence the dynamics in our temperature range. Based on these assumptions, the resonance width for an oscillator immersed in a fluid can be expressed as \cite{J.Low.Temp.Phys.146.537.Blaauwgeers}:
\begin{equation}
\Delta f = C\frac{S}{2 V\rho_\mathrm{beam}} \sqrt{\frac{\eta\rho_\mathrm{n} f}{\pi}} \left(\frac{f}{f_1}\right)^2.
\label{eqn: DeltaF}
\end{equation}
						
Figure~\ref{fig: LiquidHe4}(b) shows as a function of temperature the measured change in the resonance frequency, presented as the square of the ratio of the resonance frequency in vacuum, $f_{1}$  to that measured in the liquid, $f$ (as expressed in equation~(\ref{eqn: FreqRatio})). The rapid increase in the frequency $f$ below the superfluid transition temperature arises almost entirely from the reduction in the liquid viscously ``clamped” to the beam as a consequence of the falling normal fluid fraction. The red solid line is the least-square fit of equation~(\ref{eqn: FreqRatio}) using the parameters for liquid $^4$He tabulated by Donnelly and Barenghi \cite{JourPhysChemRefData.27.1217.Donnelly} and yields the following values for the geometric parameters: $\beta=1.18 \pm 0.02$ and $B=1.19 \pm 0.01$. The measured data are in excellent agreement with the model over the whole temperature range. Similar results have been observed for our beams with shorter lengths and higher resonant frequencies.

Figure~\ref{fig: LiquidHe4}(c) shows the measured resonance width of the beam $\Delta f$ in liquid helium as function of temperature. We found that the Stokes' drag model fits the measured data excellently until temperature of 1.7\,K below which the model starts to deviate from our measurements. The solid red curve corresponds to the least-square fit from equation~(\ref{eqn: DeltaF}) for the data between 4.2\,K and 1.7\,K with the relevant liquid $^4$He parameters from Donnelly and Barenghi \cite{JourPhysChemRefData.27.1217.Donnelly}. The least-square fit yields the geometric parameter $C=2.62\pm0.06$. The dashed red line shows the expected behaviour for the Stokes` model and highlights the presence of excess damping detected by the beam below 1.7\,K. It is clear that the intrinsic damping of this beam plays no role in the observed discrepancy since it was nearly \textit{an order of magnitude smaller} in vacuum measurements. Several effects might contribute to this difference: (\textit{i}) we could be generating quantum vorticity or exciting Kelvin waves on existing vortices attached to the beam; (\textit{ii}) it is possible that we start to observe the onset of transition from the hydrodynamic to ballistic regime in the liquid phase; (\textit{iii}) finally, we may be seeing the effect of acoustic emission.
	
Regarding the effect of turbulence, we would expect the nucleation of quantum turbulence to be triggered only at velocities an order of magnitude higher \cite{PNAS.111.4699.Vinen}. However, this has not hitherto been studied with such small objects and high frequencies. As to the possibility that we are seeing transition to the ballistic regime, unfortunately temperatures below 0.8\,K needed to observe the $T^4$ dependence expected for the phonon damping in the ballistic regime, are not accessible with the current experimental apparatus. It is also unlikely that acoustic damping is solely responsible for the discrepancy observed since some high frequency beams exhibit lower measured damping. However, a standing acoustic resonance for a particular beam and cell cannot be ruled out as the reason for the higher-than-expected dissipation at a certain temperature. Since various NEMS samples have slightly different discrepancies, it is clear that systematic measurements of several beams of various length, and consequently frequencies, down to much lower temperatures are needed to determine the actual nature of these discrepancies.
			
\section*{Discussion}
\label{sec: Discussion}

It is instructive to compare the temperature dependence of the resonance frequency of our NEMS with other devices which have been successfully used to probe the properties of superfluid helium. Using the two-fluid model we can make a reasonable prediction of which beam-like devices with a characteristic size of cross-section $d^2$, density $\rho_\mathrm{beam}$ and the resonance frequency $f$ will have the highest frequency sensitivity to the temperature-dependent changes in superfluid. First, to ease the analysis we rewrite equation~(\ref{eqn: FreqRatio}) as:
\begin{equation}
\label{eqn: FreqSens}
\left(\dfrac{f_1}{f}\right)^2 - 1 = \frac{1}{\rho_\mathrm{beam}} \left( \beta\rho + B \sqrt{\frac{\eta\rho_\mathrm{n}}{\pi}} \frac{4}{d\sqrt{f}}\right).
\end{equation}
Then, setting the geometrical coefficients $B$ and $\beta$ to be on the order of unity for all resonators, and taking the right hand side (RHS) terms one by one, it is immediately clear from the leading coefficient  that beams with lower material densities would be expected to be more sensitive. Quartz tuning forks ($\rho_\mathrm{Q}=2690$\,kg\,m$^{-3}$) and aluminium beams ($\rho_\mathrm{Al}=2700$\,kg\,m$^{-3}$) are clearly favoured over denser NbTi wires ($\rho_\mathrm{NbTi}=6500$\,kg\,m$^{-3}$). The first term on the RHS corresponds to the backflow around the resonator, has only weak temperature dependence and can be neglected in the discussion. The second term, describing viscous clamping, predicts that an object with the smallest cross-section and the lowest frequency will yield the highest frequency sensitivity. Putting this all together we find that of the devices we have tested,   the most sensitive should be the smallest diameter vibrating wire resonator manufactured from NbTi ($d = 0.9$\,$\mu$m, $f\approx10^3$) closely followed by the aluminium 50\,$\mu$m long beam ($d = 0.18$\,$\mu$m, $f\approx10^6$).  Despite being manufactured from a denser material, the vibrating wire resonator should show a sensitivity approximately two times better than our 50\,$\mu$m long beam owing to its submicron size and very low frequency. Finally the thinnest available tuning fork ($d = 25$\,$\mu$m, $f\approx10^4$) comes in at a sensitivity $\approx2.5$ times worse than our 50\,$\mu$m aluminium beam.

\begin{figure}
\includegraphics[width=0.9\linewidth]{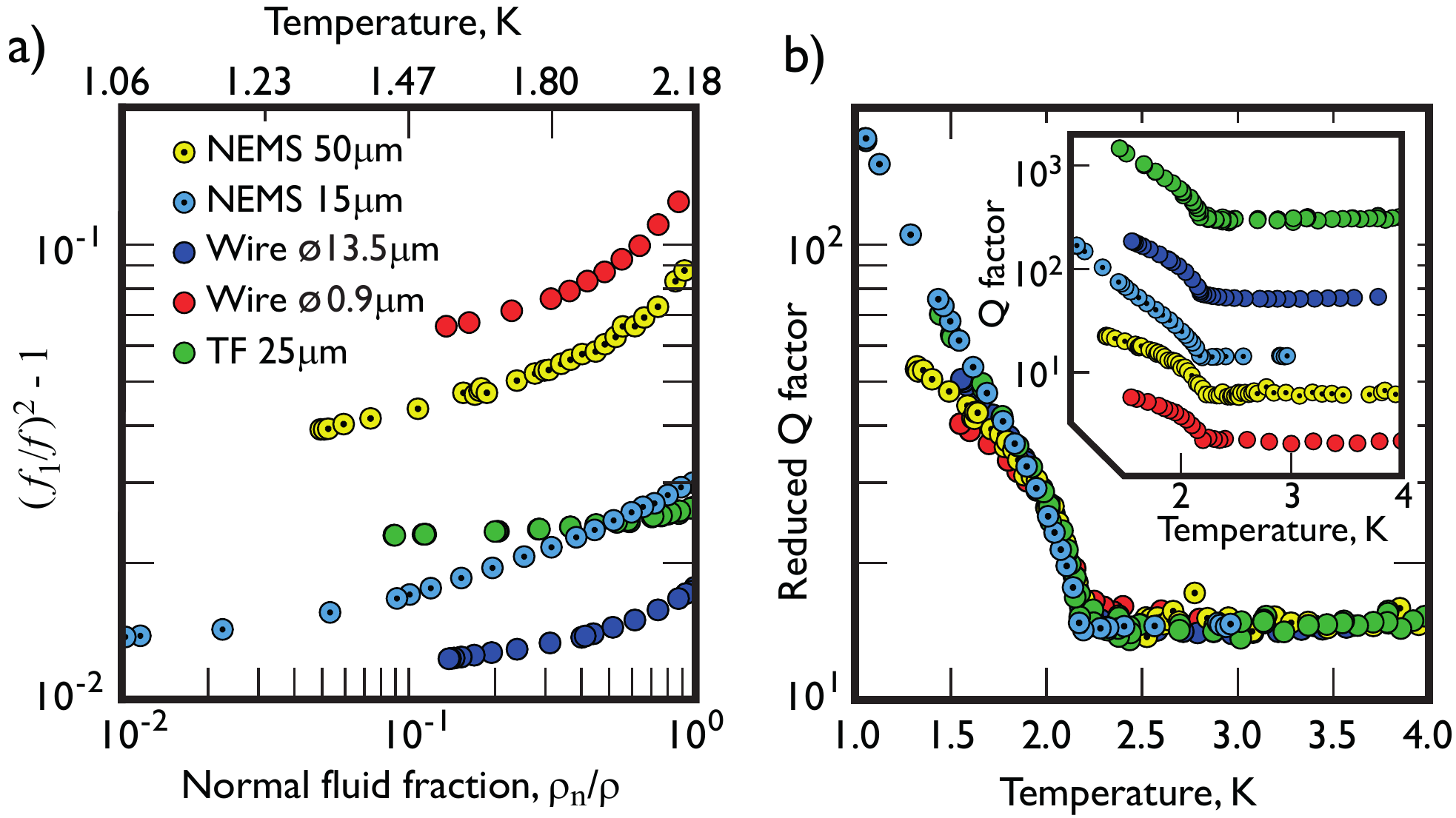}
\caption{A comparison of the frequency response of the NEMS oscillators with superconducting NbTi vibrating wires \cite{LeMiere2013,marktheodorenoble2015} and quartz tuning-fork \cite{JLowTempPhys.184.1080.Bradley} operated in the superfluid. \textbf{(a)} The frequency shift as a function of temperature. The lightness and small geometrical sizes of the NEMS resonators give a higher sensitivity to the normal fluid component of the $^4$He than the other commonly used beam-like devices. \textbf{(b)} Temperature dependence of the reduced Q-factor for various beam-like resonators. The response of the NEMS devices in the liquid is similar to that of the other devices and promises  excellent prospects for their use as detectors in quantum fluids. \textbf{Insert:} The temperature dependencies of the $Q$-factors for all resonators around the superfluid transition. Since the $Q$-factors are determined by the conventional Stokes drag, they are almost independent of temperature above the lambda-point transition $T_{\lambda}=2.178$\,K, but begin to increase significantly with the reduction of the temperature below $T_{\lambda}$ from the reduction in the density of the normal component.}
\label{fig: Comparison}
\end{figure}

Figure \ref{fig: Comparison}(a) shows the fractional frequency change for these NEMS, as a function of the normal fluid fraction $\rho_n/\rho$  from the superfluid transition down to 1.1\,K. It is clear that our 50\,$\mu$m and 15\,$\mu$m NEMS show a significant fractional frequency change over the whole temperature range, and are comparable to the smallest vibrating wire resonator and much more sensitive than a tuning fork. It is also worth noting that our NEMS do not show a significant saturation down to the lowest measured temperatures.
						
While a good frequency sensitivity is important for identifying the best resonator, a high $Q$-factor is also essential for a high signal-to-noise required for measurements. Using similar two-fluid-model arguments to those used above we can deduce from equation~(\ref{eqn: DeltaF}) that the $Q$-factor can be written as:
\begin{equation}
Q = 2C^* \sqrt{\frac{\pi}{\eta\rho_\mathrm{n}}} \left(\frac{f_1}{f}\right)^2 \sqrt{f} d \rho_\mathrm{beam},
\label{eqn: Qfactor}
\end{equation}
where we can see that the thickest, densest and highest frequency beams will have the best quality factors. This of course is a necessary trade off since these factors tend to reduce the frequency sensitivity.

The insert of figure \ref{fig: Comparison}(b) presents the $Q$-factors of various resonators in liquid helium as a function of temperature. $Q$-factors at the lowest temperatures vary by almost three orders of magnitude; the superconducting NbTi wire with a 0.9\,$\mu$m diameter and 25\,$\mu$m thick quartz tuning fork show the lowest and highest $Q$-factors, respectively. The figure insert supports our conclusions, with the NEMS devices showing much better $Q$-factors than the NbTi 0.9\,$\mu$m diameter wire, owing to their higher resonance frequencies.
			
Interestingly, the equation also suggests that the temperature dependence of the reduced $Q$-factor, $Q/(\sqrt{f_1}d\rho_\mathrm{beam})$, should be almost identical for all resonators, provided the in-liquid and vacuum resonance frequencies are similar, which is the case to within 10\% for all the data presented here. We have used the $C^*$ coefficient as a scaling parameter to normalise all of the available data and reproduce the reduced quality factor in figure~\ref{fig: Comparison}(b). Despite the very different geometries of the resonators studied, the value of $C^*$ only varies by a factor of 4 (from 0.35 to 1.4), for all the resonators. In the figure it is clear that the reduced $Q$-factors of the 15\,$\mu$m beam with resonance frequency of 8\,MHz and 25\,$\mu$m tuning fork below $\sim$1.6\,K increases more rapidly with decreasing temperature than those of the others. This would indicate that acoustic emission, which should increase very rapidly with the resonant frequency \cite{PhysRevB.85.014501.Bradley}, may not become the dominant damping for NEMS even at these low temperatures. Overall our results demonstrate that the use of NEMS beams as local superfluid temperature probes has a very exciting prospect since tuning forks and vibrating wires already work extremely well at low temperatures.									
			
The absence of any saturation of the $Q$-factor of the  high frequency beam down to $1.2\,\mathrm{K}$ warrants extending measurements down to dilution fridge temperatures. The device should enter the ballistic regime below 0.8\,K with the $Q$-factor showing a $T^4$ temperature dependence, which would then reveal whether other mechanisms such as intrinsic damping in the beam or acoustic emission become significant. Furthermore, low temperatures and correspondingly lower damping should allow us to investigate the nucleation of quantum turbulence as inferred from the force-velocity dependence of "beam + quantum fluid" system. Possibly single-vortex sensitivity may eventually become achievable using NEMS owing to the well-defined geometry, high sensitivity and small size.

Armed with the knowledge that we can run these devices in superfluid $^4$He, we can now presume that they will also run in superfluid $^3$He. We know $^3\mathrm{He}$ is the best medium to cool electronic devices to the lowest temperatures \cite{Nat.Commun.10455.Bradley}. Here the liquid temperatures can be taken routinely to temperatures below 100\,$\mu$K. This would have the immediate advantages that, first, the intrinsic $Q$-factor (inverse damping) of vibrating wire resonators immersed in superfluid $^3$He usually exceeds the $Q$-factor measured in vacuum at low resonator velocities (because the liquid $^3$He cools the material of the resonator to lower temperatures than can usually be achieved by cooling through the leads in a vacuum). Secondly, the small dimensions of the beam should make it an extremely sensitive detector of ambient quasiparticles and could lead to high resolution visualization of quantum turbulence or other topological defects present in $^3$He-B. Thirdly, the coherence length of superfluid $^3$He-B at zero pressure is 80\,nm and is of the same order as the beam dimensions, thus opening another regime to probe superfluid $^3$He-B.

However, the wider advantage would be that at these temperatures, as pointed out in the introduction, we are at the point when the low frequency nanobeams are on the brink of reaching their quantum ground state, especially if we choose a beam with somewhat higher frequency, say $\sim$10\,MHz. The challenge of performing such experiments in $^3$He is that they would require both small magnetic fields and minute power dissipation. To achieve the ultimate intrinsic $Q$-factors for the resonators, the highest sensitivity to the liquid and the necessary low dissipation would require that the aluminium is in the superconducting state. This requires an ambient field below 10\,mT, which makes the excitation and detection of the beam motion rather demanding (bearing in mind that the results reported here were made in a 5\,T field). However, this is within the capabilities of a SQUID-based voltmeter \cite{JLowTempPhys.119.703.Bradley}. Additionally, recently developed superconducting diamond beams with a critical field of several Tesla \cite{Carbon.72.100.Bautze} could be employed, or measurements utilising a microwave reflection technique that drive and detect beam motion electrostatically\cite{Nano.Lett.10.4884.Sulkko} could be adapted. We would also have to take precautions to reduce the ambient noise, but we are confident that this is possible in the ultra-quiet environments of our microkelvin cryostats \cite{JLowTempPhys.114.547.Cousins}.  

\section*{Methods}

\subsection*{Fabrication and Experimental Setup}
\label{sec: Fabrication}

The NEMS resonators used in our experiments are formed on a silicon substrate by standard nanofabrication methods: electron-beam lithography, metal deposition and reactive ion etching. Our manufacturing technique allows the creation of doubly-clamped aluminium resonators over a broad range of lengths ($l$) from 0.5\,$\mu$m up to 500\,$\mu$m. All beams have a lithographically-defined width ($w$) and thickness ($h$) with dimensions of $\approx$0.1\,$\mu$m and are clamped at both ends to a wider film of the same thickness. The beam is suspended above the substrate by 2\,$\mu$m. A scanning-electron microscope image of a typical aluminium NEMS beam is shown in figure~\ref{fig: VacuumMeasurements}(a) together with the principle measurement schematics.
	
The NEMS are mounted on a chip within a small container inside the cryogenic system and connected by two coaxial cables to a network analyser.  This setup allows the measurement of the transmission through the beam, which is characterised by a magnetomotive measurement scheme \cite{ApplPhysLett.81.2253.Ekinci}. The 5\,T magnetic field $B$ necessary for the measurements is provided by a superconducting solenoid surrounding the sample. The beam orientation is such that the Lorentz force $F_L\sim I B l$, a result of an AC current $I$ flowing through the beam, induces in–plane vibrations of the beam, with the displacement parallel to the chip surface. The transmitted signal is amplified by a low-noise high-frequency amplifier mounted at room temperature and then directed to the corresponding port of the network analyser.
		
All measurements on the beams were conducted in a $^4$He evaporation cryostat operating over the temperature range from 4.2\,K down to a base temperature of $\sim$1.2\,K. After the initial characterization of a beam in vacuum at 4.2\,K, the cell was slowly filled with helium gas while the beam resonance was monitored and followed. At pressures greater than the saturated vapour pressure at 4.2\,K, $^4$He condenses and the cell fills with a liquid. We used the damping of a vibrating wire resonator located above the beam to monitor the helium level and to confirm that the beam was indeed fully submerged. The temperature of the cryostat was monitored and controlled both by a calibrated RuO$_2$ resistor and from a measurement of the $^4$He saturated vapour pressure \cite{JourPhysChemRefData.27.1217.Donnelly}. The $^4$He saturated vapour pressure was measured at room temperature just outside the cryostat on the pumping line. Both temperature measurement methods agreed to within 20\,mK.

\section*{Data and software availability}
All data used in this paper are available at http://dx.doi.org/10.17635/lancaster/researchdata/139, including descriptions of the data sets.

\section*{Acknowledgements}
We thank S.\,M.~Holt, A.~Stokes, and M.\,G.~Ward for excellent technical support. This research was supported by the UK EPSRC Grants No. EP/L000016/1, No. EP/I028285/1 and No. EP/K01675X/1. Yu.\,A.\,P also acknowledges support from the the Royal Society Grant No. WM110105. J.\,R.\,P. acknowledges support of the People Programme (Marie Curie Actions) of the European FP7 Programme under REA grant agreement 618450.

\section*{Contributions}
R.G., S.K., M.S. and Yu.A.P. designed, fabricated and packaged the beams. D.I.B., A.M.G., S.K., M.T.N., Yu.A.P., J.R.P., M.P., M.S., R.S., V.T. and D.E.Z. developed custom measurement instrumentation and methods. S.K., M.P., M.S. and T.W. performed measurements. R.P.H., S.K., Yu.A.P., G.R.P, M.S. and V.T. performed calculations and analysis. S.K. and V.T. drafted the manuscript. All authors discussed the results and implications, and commented on the manuscript at all stages.

\section*{Competing financial interests}
The author(s) declare no competing financial interests.
	
%\bibliographystyle{nature}
%\bibliography{beaminsuperfluid}

\newpage
	
\end{document}